\PassOptionsToPackage{dvipsnames}{xcolor}
\documentclass[runningheads]{llncs}

\usepackage[dvipsnames]{xcolor}
\usepackage{algorithm}
\usepackage{algpseudocode}
\usepackage{amssymb,amsmath}
\usepackage[inline]{enumitem}
\usepackage{booktabs}
\usepackage{caption}
\usepackage{graphicx}
\usepackage{listings}
\usepackage{multirow}
\usepackage{tikz}
\usepackage{colortbl}
\usepackage{url}
\usepackage{xspace}
\usepackage[hidelinks]{hyperref}
\usepackage[capitalise,noabbrev]{cleveref}
\usepackage{tabularx}
\usepackage{subcaption}
\usepackage{pifont}
\usepackage{underscore}
\usepackage{float}
\usepackage{wrapfig}

\setlist[enumerate]{label = (\roman*)}

\renewcommand{\paragraph}[1]{\smallskip\noindent\textbf{#1.}\xspace}
\newcommand{\parbold}[1]{\smallskip\noindent\textbf{#1.}\xspace}

\algrenewcommand\algorithmicindent{1.0em}
\let\emptyset\varnothing

\newcolumntype{L}{>{\raggedright\arraybackslash}X}
\newcolumntype{R}{>{\raggedleft\arraybackslash}X}
\newcolumntype{C}{>{\centering\arraybackslash}X}

\newcommand{\syscomb}{SysComb\xspace}

\colorlet{punct}{red!60!black}
\definecolor{delim}{RGB}{20,105,176}
\definecolor{types}{RGB}{255,0,0}
\colorlet{comp}{magenta!60!black}
\colorlet{numb}{magenta!60!black}

\lstdefinelanguage{myc}{
	xleftmargin=1em,
	basicstyle=\ttfamily,
	keywords=[6]{unsigned, int, long, return, if, for, while, include, define, NULL},
	keywordstyle=[1]\color{black}\textbf,
  morecomment=[s][\color{NavyBlue}]{/*}{*/},
  morecomment=[s][\color{OliveGreen}]{"}{"},
	extendedchars=true,
	numbers=left,
	numberstyle=\scriptsize,
	stepnumber=1,
	numbersep=4pt,
	showstringspaces=false,
	showspaces=false,
	breaklines=true,
	escapechar=`,
	frame=lines,
	literate=
	*{:}{{{\color{punct}{:}}}}{1}
	{,}{{{\color{punct}{,}}}}{1}
	{=}{{{\color{comp}{=}}}}{1}
	{<}{{{\color{comp}{<}}}}{1}
	{>}{{{\color{comp}{>}}}}{1},
}

\lstdefinelanguage{mytxt}{
	xleftmargin=1em,
	basicstyle=\ttfamily,
	keywordstyle=[1]\color{black}\textbf,
  morecomment=[s][\color{blue}]{/*}{*/},
	extendedchars=true,
	numbers=left,
	numberstyle=\scriptsize,
	stepnumber=1,
	numbersep=4pt,
	showstringspaces=false,
	showspaces=false,
	breaklines=true,
	escapechar=`,
	frame=lines,
	literate=
	*{:}{{{\color{punct}{:}}}}{1}
	{,}{{{\color{punct}{,}}}}{1}
	{\{}{{{\color{delim}{\{}}}}{1}
	{\}}{{{\color{delim}{\}}}}}{1},
}

\lstdefinelanguage{myjson}{
	xleftmargin=2em,
	basicstyle=\ttfamily,
	extendedchars=true,
	numbers=left,
	numberstyle=\scriptsize,
	stepnumber=1,
	numbersep=8pt,
	showstringspaces=false,
	showspaces=false,
	breaklines=true,
	frame=lines,
	literate=
	*{0}{{{\color{numb}0}}}{1}
	{1}{{{\color{numb}1}}}{1}
	{e1}{{{e1}}}{2}
	{:}{{{\color{punct}{:}}}}{1}
	{,}{{{\color{punct}{,}}}}{1}
	{\{}{{{\color{delim}{\{}}}}{1}
	{\}}{{{\color{delim}{\}}}}}{1}
	{[}{{{\color{delim}{[}}}}{1}
	{]}{{{\color{delim}{]}}}}{1},
}

\newcommand*\emptycirc[1][0.8ex]{\tikz[baseline=-1.5pt]\draw[thick] (0,0) circle (#1);}
\newcommand*\halfcirc[1][0.8ex]{%
  \begin{tikzpicture}[baseline=-1.5pt]
    \draw[fill] (0,0)-- (90:#1) arc (90:270:#1) -- cycle ;
    \draw[thick] (0,0) circle (#1);
  \end{tikzpicture}}
\newcommand*\fullcirc[1][0.8ex]{\tikz[baseline=-1.5pt]\filldraw (0, 0)
  circle (#1);}

\newcommand{\objtemp}{\textbf{O1}}
\newcommand{\objnoac}{\textbf{O2}}
\newcommand{\objnkc}{\textbf{O3}}
\newcommand{\objnud}{\textbf{O4}}
\newcommand{\objdync}{\textbf{O5}}
\newcommand{\objsyncr}{\textbf{O6}}

\begin{document}
\title{\syscomb: Fine-Grained Transparent System Call Filtering for Attack Surface Reduction}
\titlerunning{\syscomb}

\author{Matthew Rossi \and Marco Abbadini \and Michele Beretta \and \\ Dario Facchinetti \and Stefano Paraboschi}
\authorrunning{M. Rossi, M. Abbadini, M. Beretta, D. Facchinetti, S. Paraboschi}
\institute{Università degli Studi di Bergamo, Bergamo, Italy \\
\email{name.surname@unibg.it}}

\maketitle

\begin{abstract}
  Restricting the system calls available to applications shrinks the kernel's
  attack surface and greatly mitigates the impact of compromised programs.
  Recent approaches showcase techniques to generate system call filters,
  however, all existing solutions require either kernel or application
  modifications to activate them at runtime.  This is intrusive, error-prone,
  and often impractical, especially when the code is maintained by external
  parties.  This paper presents \emph{\syscomb}, a novel eBPF-based solution to
  enforce temporally-specialized system call filters based on the application
  state, without requiring any modification to the application or the
  kernel code, and thus addressing the above limitations. Moreover, \syscomb{}
  lets the developer choose between two distinct enforcement strategies:
  \textit{seccomp-like}, ensuring no new privileges are gained after a state
  transition is performed, and \textit{least-privilege}, which applies to each
  state the most restrictive filter. We evaluated \syscomb{} using widely used
  software, showcasing accurate state-aware system call filtering and an overhead
  comparable to built-in kernel solutions, demonstrating the practicality of
  our approach.
\end{abstract}

\section{Introduction}\label{sec:intro}

Restricting the interface between the kernel and userland is essential for
safeguarding the operating system integrity. This approach serves a dual
purpose: it reduces the kernel's attack surface by
denying access to potentially vulnerable code paths~\cite{formal-syscall-interposition},
including new kernel functions that are more susceptible to
vulnerabilities~\cite{lock-in-pop}, and it limits the capabilities of
compromised applications~\cite{seminal-systrace}.
The significance of this defense-in-depth practice is evident in its
widespread adoption across critical applications such as web browsers, service
managers, and container runtimes, which restrict the functions available to
applications before their execution (e.g.,~\cite{chrome-os-seccomp,systemd,docker-seccomp}).

In Linux systems, the restriction of the available system calls (or \emph{syscalls}) is achieved using
\emph{seccomp}~\cite{seccomp}, which defines the set of allowed system calls through the
specification of a \emph{filter}.
The effectiveness of this solution strictly depends on the granularity of the
filter~\cite{seccomp-generation}. Ideally, an application should only be granted the
system calls necessary for its current (and future) operations. This requires
transitioning to a more restrictive filter whenever feasible,
or even partitioning the application functionality into
distinct threads each restricted by a dedicated, tight filter. However, in practice, most
production applications adopt an overly permissive filter at startup
and rarely modify it afterwards~\cite{seccomp-usability}. This is due to multiple reasons,
such as
\begin{enumerate*}
\item limited knowledge of the invoked system calls,
\item effort required to restructure applications to support multiple filters,
and
\item prioritization of rapid development and deployment over security.
\end{enumerate*}

Recognizing these challenges, substantial research effort has been dedicated to
generate seccomp-compatible filters using both static and dynamic analysis
techniques~\cite{temporal-specialization,confine,c2c,chestnut,b-side,syspart,sysfilter}.
Some influential works~\cite{speaker,temporal-specialization,syspart} pushed the concept even further with the
generation of two system call filters for the same server application, one for the
\emph{initialization} phase and another for the \emph{serving} phase,
demonstrating that the security of server applications can greatly benefit from
the enforcement of distinct system call sets. While these
works paved the way to the concept of \emph{temporal specialization} of system
call filters, the effective and efficient activation of these filters
at runtime still poses a significant challenge.
Indeed, existing solutions exhibit one or more of the following limitations:~\begin{enumerate*}
\item they require application modifications to explicitly
  trigger filter transitions,
\item they need to maintain an alternative patched version of the kernel with
  ad hoc features,
or
\item they delegate the enforcement of filters to a trusted user-space
  component, which represents an additional attack
  surface.
\end{enumerate*}

In this paper, we propose a novel solution to automatically identify and
transparently activate specialized system call filters based on the application
execution state. We also refine the concept of temporal specialization, enabling
the enforcement of per-thread system call filters.
To the best of our knowledge, this is the first proposal to achieve this result
without requiring application changes, kernel patches, and delegation to
user-space components. Our solution operates within the kernel boundaries
by leveraging the eBPF subsystem. This allows us to automatically detect
application state transitions at runtime using user-space probes,
observe all system call events, and enforce security decisions.
The proposed technique is flexible and supports two strategies:
\begin{enumerate*}
\item \emph{seccomp-like}, which restricts application threads to progressively
  more restrictive filters, and
\item \emph{least-privilege}, which associates each application state with the
  most restrictive filter possible.
\end{enumerate*}
Additionally, \syscomb automatically validates application state transitions at
runtime, enabling early detection and blocking of unexpected execution
patterns.
We evaluate \syscomb on widely used applications. From a security perspective,
we demonstrate that our approach identifies fine-grained, state-aware filters
that more effectively mitigate kernel vulnerability exploitation, and
blocks a larger number of shellcodes when compared to solutions based on a single filter.
From a performance standpoint, we show that SysComb
introduces an overhead comparable to seccomp.

\section{Background}\label{sec:background}

This section overviews the use of seccomp for system call filtering
and provides an essential background on eBPF, a foundational
technology used by \syscomb.

\paragraph{Seccomp}
Seccomp BPF~\cite{seccomp}, or simply \emph{seccomp}, is a Linux kernel security
mechanism that allows a process to define a \emph{syscall filter}. This filter
is a stateless program, evaluated by the kernel upon syscall invocation,
that is primarily
used to implement an allow list, permitting the execution of a system call only when its
identifier (and occasionally its primitive arguments) matches the specified
rules.
Critically, once a filter is set, it is inherited by all child threads and
processes, and it can only be tightened, never relaxed. To install a filter, an
application process must configure it and then load it into the kernel. There
are two options for filter setup:
\begin{enumerate*}
\item directly write it with a sequence of Berkeley Packet Filters
  rules and then activate it invoking \texttt{seccomp} syscall, or
\item use a platform-independent high-level library like
  \texttt{libseccomp}.
\end{enumerate*}
In either case, it is necessary to modify the application, potentially
requiring significant architectural restructuring~\cite{chromium-renderer-main}.
To reduce the effort of developers, several solutions permit
to explicitly declare the system calls available to an application
only once, before it is started (e.g.,~\cite{chrome-os-seccomp,docker-seccomp,kubernetes-seccomp}).
Although pragmatic, this approach inevitably reduces effectiveness, as complex
software needs many system calls, and there is a concrete security risk when all
of them are available to every internal component.

\begin{figure}[!t]
  \begin{center}
    \includegraphics[width=0.65\columnwidth]{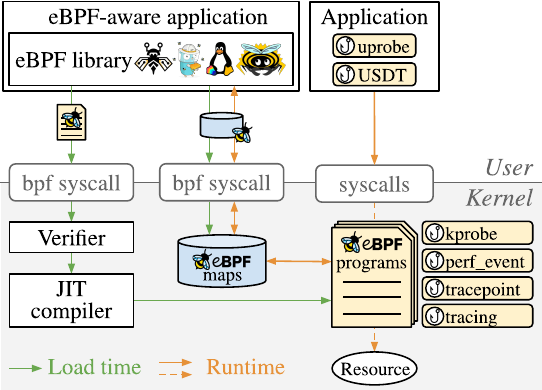}
  \end{center}
  \caption{Overview of the eBPF architecture.}\label{fig:ebpf-overview}
\end{figure}

\parbold{eBPF}
\emph{eBPF}~\cite{ebpf-lwn} is a technology designed to securely and dynamically extend
the operating system kernel's functionality without the need of source code
modifications or loading new kernel modules. Fundamentally, it enables user-defined
\emph{eBPF programs} to execute within an in-kernel abstract Virtual Machine.
These programs are triggered when the execution flow reaches designated
\emph{hook points}, which include pre-defined locations such as network
sockets and kernel trace points, and arbitrary kernel and
user-space functions when attachment is performed using \emph{kprobes} and
\emph{uprobes}, which are enabled by default in major Linux distribution. This design permits the non-intrusive inspection
of arguments and return values, and for certain hook points it also allows programs
to override how the kernel reacts to events. To manage state persistently across
invocations, programs use shared data structures known as \emph{eBPF maps} (e.g., hash
tables, arrays).
Importantly, to mitigate the risk of kernel instability, every program undergoes a
rigorous verification process prior to loading which ensures fundamental
safety properties, such as guaranteed termination, prevention of out-of-bounds
memory access, and adherence to system-defined resource limits.
\Cref{fig:ebpf-overview} presents the eBPF architecture, highlighting the
user-space component responsible for loading and interacting with eBPF programs
and maps. As shown, eBPF-aware applications rely on user-space libraries to
abstract the complexity of the low-level \texttt{bpf} system call, as
well as enable the eBPF programs to reliably target a wide range of kernel
versions without requiring recompilation.

\section{Requirements and Existing Solutions}\label{sec:req-and-solutions}

In this section we briefly formulate our objectives, clarify the
threat model, and discuss the existing
research works, highlighting their strengths and limitations.

\subsection{Objectives}\label{sec:objectives}

\begin{enumerate}
\item[\objtemp] \textbf{Fine-grained temporal filters.}
  We aim to introduce system call filters that are tailored to
  the specific execution states of each application thread, thereby
  restricting the available actions
  in potentially dangerous situations, such
  as when processing third-party data or code.
\item[\objnoac] \textbf{No application changes.}
  Temporal policies must not require the developer to refactor the application
  source code to activate the system call filters. This facilitates adoption,
  especially when third-party code needs to be restricted.
\item[\objnkc] \textbf{No kernel changes.}
  Unless patches are accepted by the kernel community and merged into
  mainstream, they must be re-applied (and potentially adapted) whenever a
  kernel update is available, introducing a considerable maintenance overhead.
  Also, errors in this code can lead to system instability.
\item[\objnud] \textbf{No user-space delegation.} Delegating security decisions
  to external monitors often adds significant delay due to
  recurrent context switches. Moreover, it always relies on trusted
  user-space components, which extend the attack surface.
\item[\objdync] \textbf{Full support of dependencies.}
  We seek to design a solution that fully supports dynamic libraries and
  execution of external binaries.
  In detail, we want to
  \begin{enumerate*}
  \item avoid over-approximation and exclude from filters all system calls
    that, although present in dynamic libraries, are not actually used by the
    application,
  and
  \item maintain the ability to manage temporal filters even inside
    dynamic libraries and external binaries.
  \end{enumerate*}
\item[\objsyncr] \textbf{Synchronous enforcement.}
  Enforcement must effectively interpose between system call invocation and
  execution, blocking forbidden system calls before dangerous actions can affect
  the system.
\end{enumerate}

\subsection{Threat Model}\label{sec:threat-model}

We consider the kernel inherently trusted but potentially affected by
vulnerabilities, while user-space applications (or specific critical
states of their execution) may be untrusted and/or potentially
vulnerable to code execution attacks. Examples include a web server
processing user requests, or a DBMS executing user-provided
queries. The identification of these application states depends on the
specific application logic, and hence is information provided by the
developer.

Equivalently to seccomp, the goal of \syscomb is to restrict the
application-kernel interface by strictly limiting the permissible
system calls available to applications, effectively reducing the
capabilities of potentially compromised software, and minimizing the
kernel's attack surface~\cite{lock-in-pop,seccomp}.
We highlight that the effectiveness of a system call filtering solution
depends on its ability to interpose between the application and the
kernel to prevent forbidden system calls. In
\syscomb, security checks are synchronous and not bypassable.

\subsection{Existing Solutions}\label{sec:ext-solutions}

We systematically analyze existing system call filtering solutions,
highlighting why they do not satisfy all our requirements.
\Cref{tab:literature-overview} summarizes the analysis.

\paragraph{User-Space System Call Filtering}
There exist several approaches that can be used to filter system
calls in user-space.
For example, \texttt{LD_PRELOAD}~\cite{ldpreload} is an environment variable
used to specify a custom shared library to be loaded before any other, which
can then implement its own version of functions and system call wrappers.
The \texttt{ptrace}~\cite{ptrace} system call allows a parent process to observe
and control the execution of another process.\ \emph{Syscall User Dispatch}
(SUD)~\cite{sud} allows an application to capture its own system calls. However,
all these methods present downsides: the scope of \mbox{\texttt{LD_PRELOAD}} is limited to
dynamic libraries, \texttt{ptrace} introduces high overhead
due to frequent context switches~\cite{gvisor-ptrace-cost}, and SUD is not
meant to be a security mechanism, as it can be easily bypassed by jumping to
non-restricted memory regions prior to system call invocation.

Another line of research studies binary rewriting techniques.
For instance, \emph{zpoline}~\cite{zpoline} rewrites every \texttt{syscall}
and \texttt{sysenter} instruction with a call to user-defined code;
\emph{bpftime}~\cite{bpftime} uses the same technique.
Although the overhead is minimal, this approach cannot trap all system calls, as
it does not support dynamic libraries.
To overcome these limitations, \emph{lazypoline}~\cite{lazypoline}
proposes the use of SUD to catch all the system calls issued by
the application, then rewrites any encountered system call instruction by
leveraging zpoline. However, being based on SUD, this approach is not suitable
to implement a security overlay.

\paragraph{Seccomp-based Solutions}
Several research works focus on the generation of
accurate system call filters, and then rely on seccomp for their
enforcement either by using application wrappers
or by injecting binary code.

Many proposals generate a single seccomp
filter for the entire application, and then apply it at startup time.
For example, \emph{sysfilter}~\cite{sysfilter} uses static analysis to generate
a function call graph from a binary, and then derives a single filter using all
reachable system calls; however, it over-approximates when dealing with dynamic
libraries, degrading the accuracy.
To improve on this aspect, \emph{Chestnut}~\cite{chestnut}
employs a two-step approach: first, it statically derives a preliminary
system call list, and then refines it through dynamic analysis.
In order to extend the approach to containers, \emph{Confine}~\cite{confine} applies dynamic analysis to
discover the list of binaries run, and then performs static
analysis on each one to derive a system call filter.
Lastly, \emph{C2C}~\cite{c2c} proposes to statically analyze the
source code of an application in conjunction with its configuration options
to identify dead code and disable all unreachable system calls: this approach is
mostly suitable to perform application debloating, is configuration-dependent,
and similarly to others is associated with low accuracy when dealing with dynamic code.
Due to their construction, all proposals in this category generate a single lax seccomp
filter, comprised of every possible system call needed by the application, and
consequently cannot support temporal policies.

On the other hand, interesting research describes novel approaches to
generate stricter filters based on the application state.
Ghavamnia et al.~\cite{temporal-specialization} (\emph{TSP}) introduce
\emph{temporal system call filters} for server applications using static
analysis on their source code.
However, this approach only supports two execution phases (initialization and
serving), and requires a patched compiler to recompile all application code,
including third-party libraries.
\emph{SysPart}~\cite{syspart} takes inspiration from this work and substantially
improves the generation of filters for binary applications
by also taking into consideration dynamic modules.
Despite the improvements, both works still rely on seccomp to install the filters,
and therefore inherit all its limitations (see \cref{sec:background}).

\begin{table}[t]
  \caption{Objectives-based comparison of recent system call filtering proposals.}\label{tab:literature-overview}
  \centering
  \begin{tabularx}{\linewidth}{l*{6}{C}}
    \multicolumn{7}{c}{O1: Temporal filters, O2: No app changes, O3: No kernel changes,}\\
    \multicolumn{7}{c}{O4: No delegation, O5: Dependency support, O6: Synchronous} \\
    \toprule
    \textbf{Proposal}                                              & \textbf{O1}  & \textbf{O2}  & \textbf{O3}  & \textbf{O4}  & \textbf{O5}  & \textbf{O6} \\
    \midrule
    \texttt{LD_PRELOAD}~\cite{ldpreload}                           & \emptycirc{} & \fullcirc{}  & \fullcirc{}  & \emptycirc{} & \halfcirc{}  & \fullcirc{} \\
    \texttt{ptrace}~\cite{ptrace}                                  & \emptycirc{} & \fullcirc{}  & \fullcirc{}  & \emptycirc{} & \halfcirc{}  & \fullcirc{} \\
    Syscall User Dispatch~\cite{sud}                               & \emptycirc{} & \emptycirc{} & \fullcirc{}  & \emptycirc{} & \emptycirc{} & \fullcirc{} \\
    zpoline~\cite{zpoline}                                         & \emptycirc{} & \fullcirc{}  & \fullcirc{}  & \emptycirc{} & \emptycirc{} & \fullcirc{} \\
    lazypoline~\cite{zpoline}                                      & \emptycirc{} & \fullcirc{}  & \fullcirc{}  & \emptycirc{} & \halfcirc{}  & \fullcirc{} \\
    bpftime~\cite{bpftime}                                         & \emptycirc{} & \fullcirc{}  & \fullcirc{}  & \emptycirc{} & \emptycirc{} & \fullcirc{} \\
    \midrule
    sysfilter~\cite{sysfilter}                                     & \emptycirc{} & \emptycirc{} & \fullcirc{}  & \fullcirc{}  & \halfcirc{}  & \fullcirc{} \\
    TSP~\cite{temporal-specialization}                             & \halfcirc{}  & \emptycirc{} & \fullcirc{}  & \fullcirc{}  & \emptycirc{} & \fullcirc{} \\
    Chestnut~\cite{chestnut}                                       & \emptycirc{} & \fullcirc{}  & \fullcirc{}  & \fullcirc{}  & \halfcirc{}  & \fullcirc{} \\    
    C2C~\cite{c2c}                                                 & \emptycirc{} & \fullcirc{}  & \fullcirc{}  & \fullcirc{}  & \halfcirc{}  & \fullcirc{} \\
    SysPart~\cite{syspart}                                         & \halfcirc{}  & \fullcirc{}  & \fullcirc{}  & \fullcirc{}  & \halfcirc{}  & \fullcirc{} \\
    Confine~\cite{confine}                                         & \emptycirc{} & \fullcirc{}  & \fullcirc{}  & \fullcirc{}  & \halfcirc{}  & \fullcirc{} \\
    \midrule
    SysXCHG~\cite{sysxchg}                                         & \emptycirc{} & \emptycirc{} & \emptycirc{} & \fullcirc{}  & \halfcirc{}  & \fullcirc{} \\
    Jia et al.~\cite{programmable-syscalls-ebpf}                   & \fullcirc{}  & \fullcirc{}  & \emptycirc{} & \fullcirc{}  & \fullcirc{}  & \fullcirc{} \\
    B-Side~\cite{b-side}                                           & \fullcirc{}  & \fullcirc{}  & \halfcirc{}  & \halfcirc{}  & \halfcirc{}  & \hspace*{0.1em}$*$         \\
    \midrule
    Optimus~\cite{optimus}                                         & \emptycirc{} & \fullcirc{}  & \fullcirc{}  & \fullcirc{}  & \halfcirc{}  & \fullcirc{} \\
    KubeRosy~\cite{kuberosy}                                       & \emptycirc{} & \fullcirc{}  & \fullcirc{}  & \fullcirc{}  & \emptycirc{} & \fullcirc{} \\
    $\mu$PolicyCraft~\cite{upolicycraft}                           & \fullcirc{}  & \fullcirc{}  & \fullcirc{}  & \emptycirc{} & \halfcirc{}  & \emptycirc{} \\
    GoLeash~\cite{goleash}                                         & \halfcirc{}  & \fullcirc{}  & \fullcirc{}  & \emptycirc{} & \halfcirc{}  & \emptycirc{} \\
    Tetragon~\cite{tetragon}, KubeArmor~\cite{kubearmor}           & \emptycirc{} & \fullcirc{}  & \fullcirc{}  & \fullcirc{}  & \fullcirc{}  & \fullcirc{} \\
    Tracee~\cite{tracee}, Falco~\cite{falco}, SysDig~\cite{sysdig} & \emptycirc{} & \fullcirc{}  & \fullcirc{}  & \emptycirc{} & \fullcirc{}  & \emptycirc{} \\
    \midrule
    \rowcolor{lightgray}Our solution                               & \fullcirc{}  & \fullcirc{}  & \fullcirc{}  & \fullcirc{}  & \fullcirc{}  & \fullcirc{} \\         
    \bottomrule
    \multicolumn{7}{c}{\emptycirc{} not supported, \halfcirc{} partially supported, \fullcirc{} supported,} \\
    \multicolumn{7}{c}{$*$ implementation-dependent} \\
  \end{tabularx}
\end{table}

\paragraph{Solutions Requiring Kernel Changes}
To overcome seccomp's limitations, several approaches propose to change
the kernel implementation.
\mbox{\emph{SysXCHG}}~\cite{sysxchg} allows a program to change its seccomp filters
every time an \texttt{execve} is performed, in accordance with the \emph{least-privilege}
principle. However, it requires to extensively modify the seccomp inheritance policy
and has to embed filters in ELF files.
Jia et al.~\cite{programmable-syscalls-ebpf} propose a modification of the
seccomp and eBPF subsystems to implement programmable system call filtering,
thereby enabling the enforcement of temporal policies. However, the proposed patch
did not gain sufficient consensus within the Linux kernel community, and was
ultimately not incorporated into the mainline kernel. Lastly,
\emph{B-Side}~\cite{b-side} focuses on static analysis of binaries to detect
states and related filters with relatively high accuracy, but does not support
filter enforcement, recommending either kernel changes or the
introduction of user-space monitors.

Although kernel modification unlocks a plethora of opportunities, it also
introduces significant drawbacks. Unmerged changes require constant
application and revision of patches across kernel updates, leading to a
considerable and costly maintenance burden.

\paragraph{eBPF-based Solutions}
A novel line of research focuses on containerized applications
and uses eBPF to improve system call
filtering~\cite{optimus,kuberosy,upolicycraft,goleash,poster-syscalls,dmng}.
\emph{Optimus}~\cite{optimus} detects the list of system calls required by an
application, while \emph{KubeRosy}~\cite{kuberosy} continuously enforces 
developer-provided seccomp filters. However, both do not support temporal policies.
\emph{$\mu$PolicyCraft}~\cite{upolicycraft} goes beyond this limitation,
allowing for the generation and enforcement of temporal policies through a
user-space monitor.
Yet, as security decisions are deferred until the monitor processes the event,
enforcement occurs only after potentially dangerous actions have already been
executed.
Focusing on Go programs, \emph{GoLeash}~\cite{goleash} generates and enforces
per-package system call filters. Although it supports temporal specialization,
it delegates the enforcement to an asynchronous user-space component.

Nowadays, many cloud-native solutions such as \emph{Falco}~\cite{falco},
\emph{KubeArmor}~\cite{kubearmor}, \emph{Sysdig}~\cite{sysdig},
\emph{Tetragon}~\cite{tetragon}, and \emph{Tracee}~\cite{tracee} use eBPF to
monitor and restrict workloads.
While KubeArmor and Tetragon provide in-kernel enforcement capabilities, the
others focus on the detection of events, leaving enforcement to user-space
modules. All of them configure a single filter for an entire container, do not
support temporal specialization, and employ block-list strategies, which are
inherently less robust than allow-list approaches. For instance, a pre-existent
filter does not block newly-introduced system calls.

\section{\syscomb Architecture}\label{sec:design}

We first present an overview of our solution by highlighting its
properties in \cref{sec:overview}, then describe in detail each
component in dedicated sections.

\subsection{Function Overview}\label{sec:overview}

In this paper, we refine the \emph{temporal specialization} concept by
associating each application thread (or process) with an execution
state, which can evolve at runtime (\objtemp). Indeed, we ensure
transitions to new states occur when the thread reaches one of the
developer-specified code locations that we refer to as \emph{transition
points}. To detect them, we set eBPF user probes in application binaries
(including external dependencies, \objdync), to trigger the execution of
eBPF programs without requiring any modification to the application source code
or binary (\objnoac). This permits to transparently track the current state for each application
thread, and evaluate the state management logic whenever a transition point is reached.

\begin{wrapfigure}{r}{0.30\textwidth}
  \centering
  \vspace{-2.4em}
  \includegraphics[width=\linewidth]{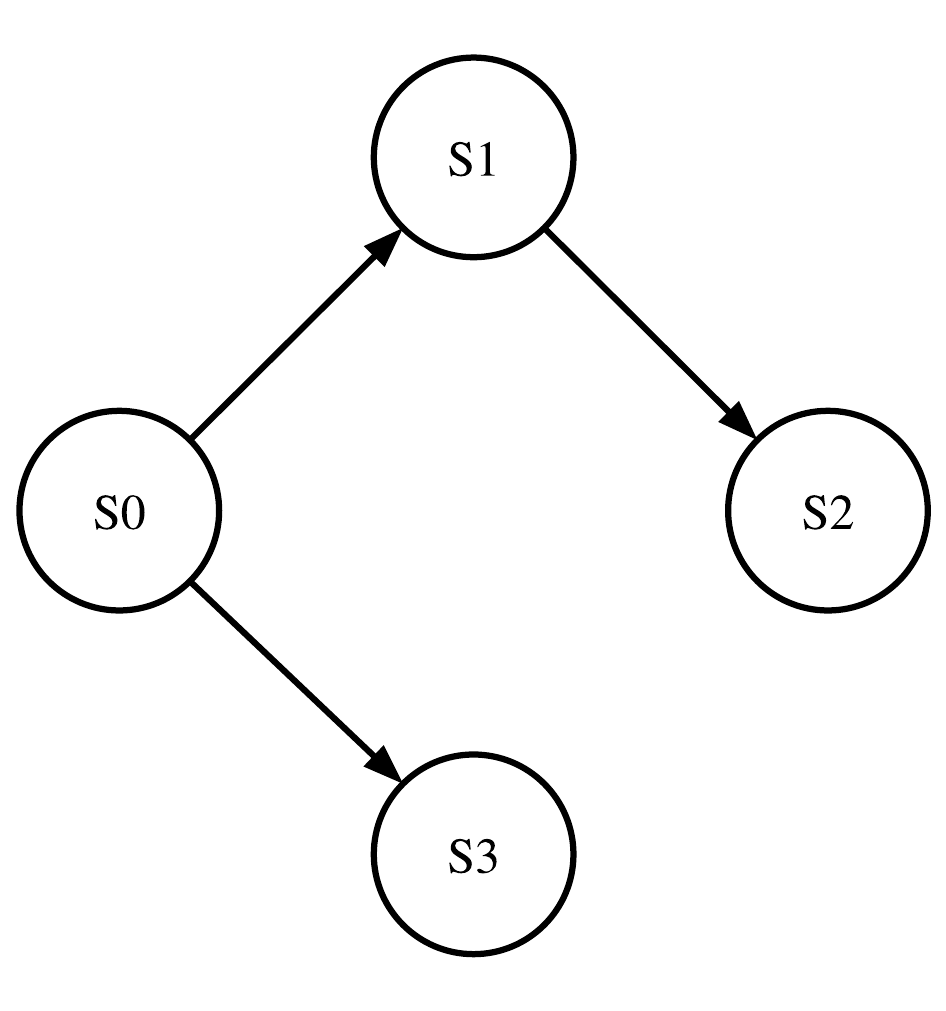}
  \vspace{-2.0em}
  \caption{State graph\vspace{-2.4em}}%
  \label{fig:visual:graph}
\end{wrapfigure}

To capture system call events and enforce security decisions, we configure eBPF
kernel probes and programs on system call handlers. During filter identification,
in \emph{tracing mode}, this setup records all system calls issued
by the application threads and associates them with the current application state.
In \emph{enforcement mode}, it detects system calls denied by the current
filter and triggers an enforcement action (i.e.,~logging the event, returning
a permission-denied error, or killing the target thread). All these operations
are seemlessly performed in kernel-space before the system call is executed
(\objnud, \objsyncr). No modifications to the kernel or the eBPF subsystem are
required; hence, our solution is readily deployable on any Linux system running
kernel v6.6 or later (\objnkc).

\begin{figure}[t]
  \centering
  \begin{subfigure}[t]{0.30\textwidth}
    \centering
    \includegraphics[width=\linewidth]{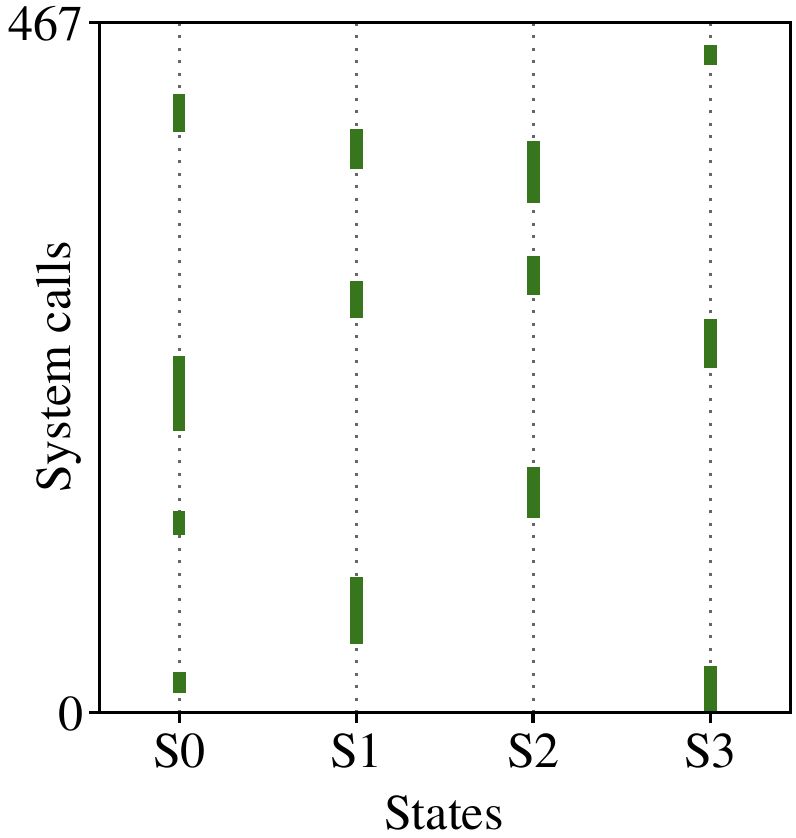}
    \caption{Least-privilege}%
    \label{fig:visual:least-privilege}
  \end{subfigure}
  \begin{subfigure}[t]{0.30\textwidth}
    \centering
    \includegraphics[width=\linewidth]{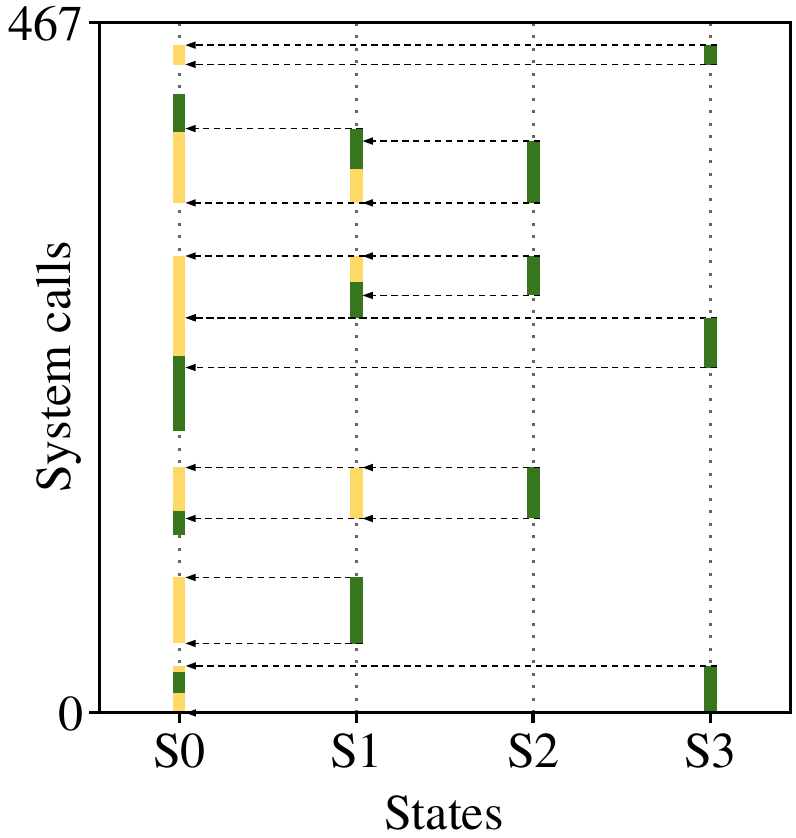}
    \caption{Backpropagation}%
    \label{fig:visual:back-propagation}
  \end{subfigure}
  \begin{subfigure}[t]{0.30\textwidth}
    \centering
    \includegraphics[width=\linewidth]{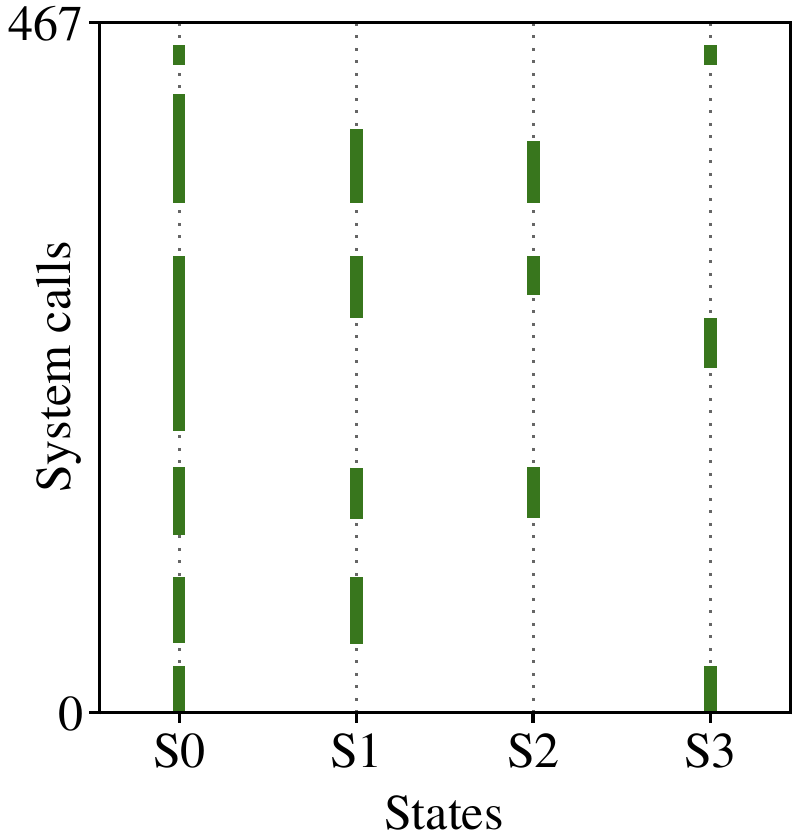}
    \caption{Seccomp-like}%
    \label{fig:visual:seccomp-like}
  \end{subfigure}
  \caption{Visualization of the system calls allowed for each
    application state in the least-privilege and seccomp-like
    strategies. The effect of flow-based backpropagation is
    highlighted with dashed arrows and the yellow areas.}%
	\label{fig:visual:syscalls}
\end{figure}

\syscomb{} supports two enforcement strategies:
\begin{enumerate*}
  \item \emph{seccomp-like} mimics the seccomp security model by always
    transitioning to more restrictive filters, but does so transparently,
  and
  \item \emph{\mbox{least-privilege}} binds each application state to
    the smallest set of system calls, yielding finer-grained filters.
\end{enumerate*}
In tracing mode, the former leverages a post-processing procedure we called \emph{flow-based
backpropagation of system calls}, which ensures that no new privileges are
gained due to a state transition, while the latter does not. However, both strategies
rely on the state management logic to record the state transitions of the
application in a directed graph. This is key to ensure an attacker hijacking the
control flow of the program cannot abuse state transitions to deliberately gain
privileges. \Cref{fig:visual:graph,fig:visual:syscalls} showcase a simple
application state graph and their associated system call filters, highlighting
the effect of the flow-based backpropagation of system calls.

\subsection{Application State Management}\label{sec:app-state}

\paragraph{Application Tracking}
To simplify the description, we call the monitoring component the \emph{tracer}
and any application thread or process being tracked a \emph{tracee}.
The tracer provides an interface to start tracking tracees that are already
running on the host, as well as new processes.
Existing threads are identified using the group of threads to trace (i.e., user,
process, or thread ID), while new processes are tracked via fork-exec pattern
and an eBPF program attached to the ELF entry point of the application binary.

To distinguish between tracees and other tasks, the tracer annotates the
identity and execution state of each tracee into an eBPF map named \emph{tracee
map}. Over the tracer's lifetime, this map is updated with new insertions and
deletions. New tracees, created with clone and fork system calls, inherit the same
properties of the parent task, while tracees that terminate are
removed. To this end, the tracer attaches eBPF programs to the kernel
raw tracepoint associated with the \texttt{sched_process_fork} and
\texttt{sched_process_exit} events, respectively.

\begin{figure}[t]
  \centering
  \includegraphics[width=0.65\linewidth]{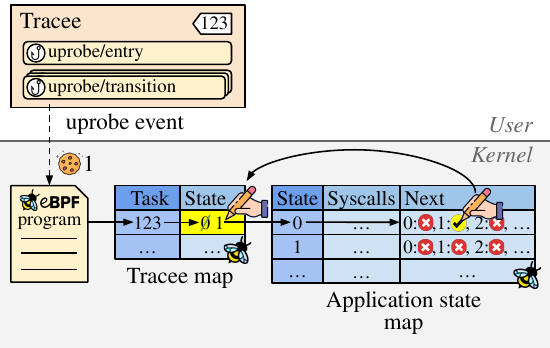}
  \caption{Application state management in tracing mode.
    \syscomb{} tracks thread state by attaching eBPF uprobes at transition points
    and builds the application state graphs.}%
  \label{fig:asm-tracing}
\end{figure}

\begin{figure}[t]
  \centering
  \includegraphics[width=0.65\linewidth]{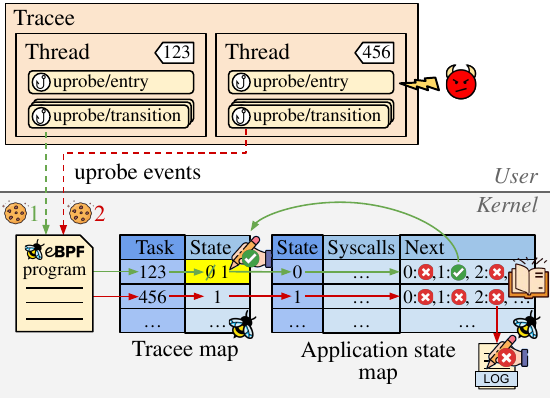}
  \caption{Application state management in enforcement mode.
    At runtime, \syscomb{} validates transitions against the state graph built
    during tracing.
  }\label{fig:asm-enforcement}
\end{figure}

\paragraph{Detection of Transitions}
To implement the
support for multiple states, we first expect the developer
to specify the transition points. These can be
located in the application, as well as its dependencies, and
can be either
\begin{enumerate*}
  \item a symbol, i.e., the name of the function,
  \item an offset from a symbol,
  or
  \item an absolute address.
\end{enumerate*}
The tracer relies on the \emph{blazesym} library~\cite{blazesym} to resolve
symbols from the \texttt{.symtab} and \texttt{.dynsym} sections of the ELF
binary, as well as from debugging symbol packages available on the system that
can be separately installed.
Then, the tracer loads into the kernel an eBPF map called \emph{application
state map}, which stores system call filters and valid transitions for each
execution state. This map is populated in tracing mode, and is read-only in
enforcement mode. Writes and reads to this eBPF map occur any time a tracee
reaches a transition point. To implement this, the tracer attaches a uprobe eBPF
program to all application transition points. As transition points can be many,
rather than performing single attachments, the kernel provides the means to
create a single eBPF program that can be attached to multiple uprobes, as well
as to specify an integer number called \emph{cookie} for each attachment point,
allowing us to unambiguously detect the transition points and consequently know
the application states they lead to.
This allows to keep the execution state in the tracee map up to date. Moreover,
by tracking states and transitions the tracer is able to build the application
state graph when in tracing mode (\cref{fig:asm-tracing}), and to validate
transitions when in enforcing mode (\cref{fig:asm-enforcement}). Indeed, during
enforcement, any transition that does not match the application state graph
is considered invalid and reported as a flow-integrity error. This permits
to detect and prevent attacks that alter the control flow of a vulnerable
application with the goal of reaching a set of filters that is more permissive
than the ones reachable from the current state.

\begin{figure}[t]
  \centering
  \includegraphics[width=0.65\linewidth]{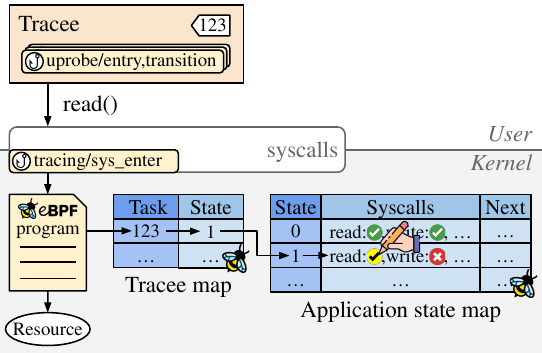}
  \caption{Tracing system calls used by each application state.}%
	\label{fig:modes:tracing}
\end{figure}

\paragraph{Detachment}
When all tracees terminate, the tracer can collect the information
recorded (if in tracing mode), and then unlink the loaded eBPF maps and
programs. To implement this, the tracer keeps an atomic counter
representing the number of tracee tasks alive. The counter is incremented
whenever a new tracee is created, and decremented whenever one terminates.
Besides natural termination, early termination of the tracer is also supported
either by sending a \texttt{SIGINT} signal to the tracer process (i.e.,~Ctrl+C)
or, only in the case of tracing, with the expiration of a maximum duration
timer.
Based on the type of tracee (i.e.,~started using an entry point vs previously
running processes), the early termination of the tracer leads to different
outcomes. In the case where an application entry point is specified, the tracer
stops all the associated tracees by sending a \texttt{SIGTERM} signal, and then
it waits for their termination. This ensures that all tracee tasks are properly
terminated, and all the system call invoked in their termination procedures are
captured. On the other hand, when the tracer monitors processes already running
on the hosts, the tracer stops tracing them without disrupting their execution.

\subsection{Identification of System Call Filters}\label{sec:filter-generation}

\paragraph{System Call Tracing}
System call events are captured by attaching an eBPF program to the \texttt{sys_enter}
kernel raw tracepoint.
As shown by the tracepoint format in \cref{lst:sys-entry-format}, this allows
the eBPF program to receive information about the system call number and its
arguments. Hence, for tracee tasks, the eBPF program retrieves the current state
from the tracee map, then uses it as an index to the application state map to
retrieve the associated system call bitmap, and finally sets the corresponding
bit based on the system call number (see \cref{fig:modes:tracing}).
Currently, the tracer records only the system call identifier. However, the
approach can be extended to track primitive-type arguments as well, similarly to
seccomp (i.e.,~without pointer dereference to avoid TOCTTOU issues).
\begin{lstlisting}[
  abovecaptionskip=0em,
  caption=Format of the raw tracepoint for system call entry.,
  float=tp,
  floatplacement=tbp,
  language=mytxt,    
  label=lst:sys-entry-format,
  keywords={id,args},
]
field:long id; offset:8; size:8; signed:1;
field:unsigned long args[6]; offset:16; size:48; signed:0;
print fmt: "NR %ld (%lx,%lx,%lx,%lx,%lx,%lx)",`\break\vspace{0.25em}`REC->id, REC->args[0], REC->args[1],`\break\vspace{0.25em}`REC->args[2], REC->args[3], REC->args[4],`\break\vspace{0.25em}`REC->args[5]
\end{lstlisting}

\paragraph{Post-processing of the Filter Set}
After the tracing terminates, the tracer starts post-processing the information
recorded to generate the filter set.
First, unused transition points are detected and prompted to the developer, as
they are not useful in the definition of the filter set. Then, if the developer
opted for the seccomp-like strategy, the tracer performs an additional step, the
\emph{flow-based backpropagation of system calls}.
This process ensures that, following a sequence of valid
transitions, the application is associated with a progressively more
restrictive filter (i.e., no new privileges are ever gained).
The implementation takes into account the potential existence of cycles in
the application state graph by retrieving the strongly connected
components in reverse topological order, merging all system calls sets
belonging to the same connected component, and backpropagating
system calls to source nodes (see \cref{alg:flow-backpropagation} in
Appendix~\ref{sec:backpropagation}). The graph is subsequently reduced by
pruning redundant states that have the same system call set as every node that
precedes them.
This simplifies the application state graph, and consequently decreases the size
of the application state map and the number of uprobes attached to the
application binary. Finally, the tracer exports the result to a JSON policy file
(exemplified in \cref{lst:policy-example}) reporting information about
\begin{enumerate*}
\item the application and entry point if available,
\item the strategy (\emph{seccomp-like} vs \emph{least-privilege}),
and
\item a list of application states comprised of state identifier,
  transition point (i.e., path, symbol name, offset, or absolute address),
  allowed system calls in human-readable format, and
  reachable states.
\end{enumerate*}

It is worth noting that, when employing the \emph{least-privilege} strategy,
the flow-based backpropagation of system calls is not performed, and hence the
identified system call filters can be significantly more
restrictive. However, valid state transitions may lead to the
activation of more permissive filters. While this may be considered a
risk, we remind that filters are not set by the application itself,
but rather by the trusted in-kernel eBPF programs according to the
security policy. Moreover, even applications affected by control-flow
hijacking attacks cannot escape system call filters.  Indeed, the
implemented control-flow integrity checks ensure that system calls can
only be exercesed based on the sequence of reachable states.

\begin{lstlisting}[
  abovecaptionskip=0em,
  caption=Seccomp-like policy generated for the Nginx web server.,
  float=tp,
  floatplacement=tbp,
  language=myjson,
  label=lst:policy-example,
]
{
  "mode": "seccomp",
  "states": [
    {
      "id": 0,
      "syscalls": [
        "read", "write", "close", "sendto",
        "listen", "getsockname", ...
      ],
      "next": [1]
    },
    {
      "id": 1,
      "probe": {
        "path": "/usr/bin/nginx",
        "symbol": "ngx_master_process_cycle"
      },
      "syscalls": [
        "read", "write", "close", "sendto",
        "getsockname", ...
      ],
      "next": []
    }
  ]
}
\end{lstlisting}

\subsection{Enforcement of System Call Filters}\label{sec:filter-enforcement}

\paragraph{Policy Loading}
In enforcement mode, the tracer expects as input a JSON policy file declaring
the filters (e.g.,~\cref{lst:policy-example}). After parsing and validating the
file format, its content is loaded into the application state map.  Note that,
while \syscomb operating in tracing mode supports the generation of such syscall
filters, they can also be produced by other tools or even written manually.

\paragraph{Enforcement Actions}
Based on the selected configuration, the tracer can
react to policy violations in three different ways:
\begin{enumerate*}
  \item log the event,
  \item inject and return an error code,
  or
  \item kill the tracee.
\end{enumerate*}
These are common actions implemented by existing system calls
filtering solutions such as seccomp~\cite{seccomp}, and can
accommodate for several application requirements.  For instance, with
logging it is possible to keep track of violation events without
altering the application flow, a useful feature when prototyping or
fine-tuning filters. On the other hand, returning an error code permits
to enforce the security restriction, and, at the same time, lets the
application recover from the error and continue its execution. This is
convenient for critical applications that need to be highly
available. Finally, killing the tracee ensures
termination after a security violation is detected.

\paragraph{Action-based Attachment}
Based on the action requested by the developer, the eBPF program
responsible to perform the security enforcement is attached to
different hook points. When the action requested is logging, the eBPF
program is directly attached to system call entry kernel
tracepoint, namely \mbox{\texttt{sys_enter}}. When the selected actions are
instead returning an error code or killing, the process is more elaborate and
requires more explanation.

\begin{lstlisting}[
  abovecaptionskip=0em,
  caption=Macro for system calls with no arguments~\cite{syscall-macros}.,
  float=tp,
  floatplacement=tbp,
  language=myc,
  label=lst:syscall-define0,
  emph={ALLOW_ERROR_INJECTION},
  emphstyle={\color{Green}\textbf},
]
#define SYSCALL_DEFINE0(sname)                            \
  SYSCALL_METADATA(_##sname, 0);                          \
  asmlinkage long sys_##sname(void);                      \
  ALLOW_ERROR_INJECTION(sys_##sname, ERRNO);              \
  asmlinkage long sys_##sname(void)
\end{lstlisting}
\begin{lstlisting}[
  abovecaptionskip=0em,
  caption=Macro for system calls with arguments~\cite{syscall-macros}.,
  float=tp,
  floatplacement=tbp,
  language=myc,
  label=lst:syscall-definex,
  emph={ALLOW_ERROR_INJECTION},
  emphstyle={\color{Green}\textbf},
]
#define SYSCALL_DEFINEx(x, sname, ...)                    \
  SYSCALL_METADATA(sname, x, __VA_ARGS__)                 \
  __SYSCALL_DEFINEx(x, sname, __VA_ARGS__)

#define __SYSCALL_DEFINEx(x, name, ...)                   \
  [...]                                                   \
  asmlinkage long                                         \
    sys##name(__MAP(x,__SC_DECL,__VA_ARGS__))             \
  [...];                                                  \
  ALLOW_ERROR_INJECTION(sys##name, ERRNO);                \
  [...]                                                   \
  static inline long                                      \
    __do_sys##name(__MAP(x,__SC_DECL,__VA_ARGS__))
\end{lstlisting}

For safety reasons, there are limitations on the hook points
where eBPF programs can alter the normal kernel execution flow, and
tracepoints (used with the logging action) are subject to these restrictions.
Fortunately, the Linux
kernel allows to inject errors and modify the return value of all
functions implementing the entry to each system call.  The complete
coverage of system calls can then be assessed by inspecting the kernel
source code and, specifically, by checking the use of the
\texttt{ALLOW_ERROR_INJECTION} macro in the declaration of the
system call functions (\cref{lst:syscall-define0,lst:syscall-definex}).%
\footnote{Similar declarations are used for compatibility system call functions.}
While this approach ensures all system calls are covered, the lack of a
single hook point for all system calls means that the tracer needs to deal
with numerous hook points (currently about 1000 different
functions).
Luckily, the kernel allows to perform this operation in a
scalable manner by creating a single eBPF program that is attached to
multiple \emph{kprobes}, using again an integer \emph{cookie} to
discriminate among the hook points, which in this case is
parameterized with system call numbers. To retrieve this number, the tracer builds a mapping based on
\begin{enumerate*}
  \item the list of error injection functions supported by the kernel,
  and
  \item the relation between the functions implementing system calls and
    their system call numbers.
\end{enumerate*}
The former is provided by the kernel through \emph{sysfs}, a pseudo
file system exporting information about various kernel
subsystems. Hence, user-space processes can read the complete list
of error injection functions 
at \texttt{\footnotesize /sys/kernel/debug/tracing/events/error_injection}.
The latter is generated during kernel compilation, and is either
available in the Linux header files, or can be downloaded from the
kernel code repository.

Independently of the action-based attachment, the tracer is able to
interpose all system calls.  Yet, not all
attachment methods are equally performant, as when a raw tracepoint is
used, the tracer loads an eBPF \emph{tracing program}, a more
modern alternative to kprobes and tracepoints, which relies on eBPF
trampolines to reduce latency (more details in \cref{sec:runtime-evaluation}).

\begin{figure}[t]
	\centering
  \includegraphics[width=0.65\linewidth]{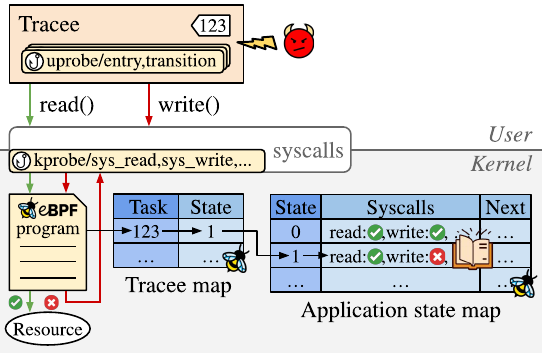}
	\caption{Enforcement of temporally-specialized system call filters. Any system
	call denied by the state map causes the eBPF program to execute the
	enforcement action (e.g., an error is returned when \texttt{write} is
	invoked).}%
	\label{fig:modes:enforcement}
\end{figure}

\paragraph{Enforcement}
When a system call is invoked, an eBPF program responsible for
enforcement is triggered. First it determines whether the task issuing
the system call is a tracee. If so, it reads the current task's
application state from the tracee map, and then retrieves the bitmap
encoding the filter associated with it.  Finally, it uses the system call
identifier, that is either given as an argument to the eBPF program,
or from the invocation of the
\mbox{\texttt{bpf_get_attach_cookie}} helper function, as an index to test
the system call bitmap. If the filter allows the system call, the eBPF program
returns and the kernel continues its normal execution path. Otherwise,
the enforcement action selected by the developer is performed.
A simplified visualization of the process is depicted in
Figure~\ref{fig:modes:enforcement}.

Note that, when sending the kill signal to the
application, the kernel still continues with the execution of the
system call until a signal handling safe point is reached.
On the other hand,
when an error is injected, the system call entry function is
overridden, and its code is not executed at all. This behavior comes
with security implications. Indeed, while killing the tracee may seem
effective, it does not prevent the system call code from being at least
partially executed, and this may result in undesired side-effects.
Hence, in the implementation we opted to always inject an error code
when a kill signal is sent to the tracee. This ensures that the
system call code is never executed, and prevents side-effects.

\subsection{Limitations}

\paragraph{Green Threads}
Some programming languages implement \emph{green threads},
i.e., lightweight threads managed by a runtime library or a virtual
machine. Green threads generally incur lower overhead than operating
system threads during a context switch, since the operation does not
involve the
kernel. Indeed, green threads are typically multiplexed into a single
operating system thread. Currently, our proposal assigns a state to
each operating system thread, hence it cannot be directly applied to
green threads as is.
However, green threads can be supported by attaching uprobes to the runtime
library or virtual machine functions responsible for their creation and
multiplexing.

\paragraph{Interpreted Programming Languages}
Not all programming languages compile to native code at build-time, as
some (e.g.,~Java, JavaScript, Python) are either interpreted or just-in-time
compiled, and execute within a runtime.
\syscomb is not directly compatible with these languages, as it relies on the
attachment of user probes to the application binary to transparently transition
the execution state.
However, since the runtime itself is always written in native code, it is
possible to attach user probes to the runtime functions, and perform state
transitions based on their arguments. The identification of the attachment
points and the assessment of the relationship between the runtime and
application operations unfortunately requires knowledge of the runtime
internals.

\paragraph{Privileged Tracer}
Contrary to seccomp, which lets unprivileged users shrink their filters,
part of \syscomb requires superuser privileges to load
eBPF programs and maps.
It is important to note that this only applies to the tracer, and that after
eBPF objects are loaded, superuser privileges are no longer required and are
automatically released before the application is executed.

\section{Discussion}\label{sec:discussion}

We discuss potential attacks and countermeasures implemented by
\syscomb.

\paragraph{Tampering with eBPF Objects}
Contrary to seccomp, which guarantees system call filter persistence
until process termination, the correct operation of our approach
depends on the integrity of eBPF programs and maps. The attacker
can attempt to tamper with filters either by
\begin{enumerate*}
  \item directly changing these eBPF objects through the \texttt{bpf} system
    call, or
  \item by indirectly triggering their removal forcefully terminating the tracer
    process that holds their references.\footnote{Like other kernel resources,
    eBPF objects are reference counted; the kernel automatically frees them when
    all references are dropped.}
\end{enumerate*}
To explain possible defenses, we discern between unprivileged and privileged
attackers. An unprivileged attacker cannot directly use the eBPF subsystem,
hence can only perform an indirect attempt. To thwart their attempts, \syscomb
pins all eBPF objects to the eBPF filesystem (\textit{bpffs})~\cite{pinning}.
Instead, to prevent direct changes by a privileged attacker, we
configure additional security mechanisms through Linux Security Modules (LSMs).
Indeed, LSMs apply to unprivileged and privileged processes, and can restrict
access to kernel resources (including eBPF objects) through mandatory access
control policies.
Currently, beside common LSMs like SELinux~\cite{selinux}
and AppArmor~\cite{apparmor}, several recent
works~\cite{kubearmor,cage4deno,enhance-wasm-sandbox,natisand} leverage
eBPF programs~\cite{ebpf-paper} to implement ad-hoc mandatory access control
solutions.

\paragraph{Control-Flow Hijack Attacks}
While both seccomp and \syscomb reduce the impact of compromised applications
by restricting the allowed system calls, neither prevent the exploitation of
user-space vulnerabilities that divert their control flow.
In \syscomb this could lead to jumping to arbitrary transition points with the
intent of installing more permissive filters. To circumvent this threat, when in
enforcing mode, we only allow the state transitions defined in the application
state graph.
Hence, an attacker cannot gain access to system calls that are unreachable
from the current state. Moreover, while not the primary goal of our proposal,
it enables coarse-grained detection of control-flow integrity violations, and
triggers enforcement actions.

\section{Effectiveness Evaluation}\label{sec:filter-gen}

To assess the effectiveness of our solution, we
\begin{enumerate*}
  \item validated the ability to trace all system calls invoked by an
        application,
  \item evaluated how the size of filters evolves during the lifecycle
        of server applications,
   and finally
  \item showcased the security benefits by analyzing the
        ability to block recent kernel vulnerabilities.
\end{enumerate*}

\subsection{Accurate System Call Detection}\label{subsect:eval-tracing}

\begin{figure*}
  \centering
  \begin{subfigure}[b]{\linewidth}
    \centering
    \includegraphics[width=0.6\linewidth]{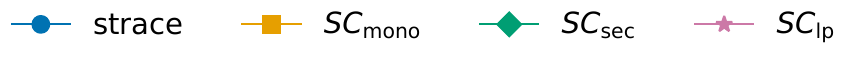}
  \end{subfigure}
  \begin{subfigure}[b]{0.49\linewidth}
    \includegraphics[width=\linewidth]{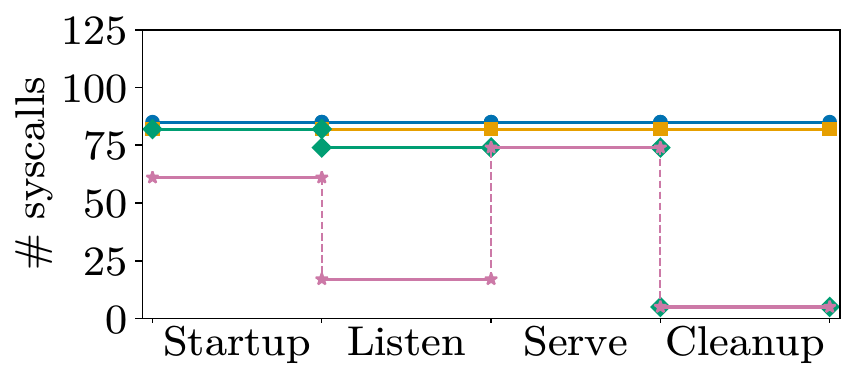}
    \caption{Apache Httpd}
  \end{subfigure}
  \begin{subfigure}[b]{0.49\linewidth}
    \includegraphics[width=\linewidth]{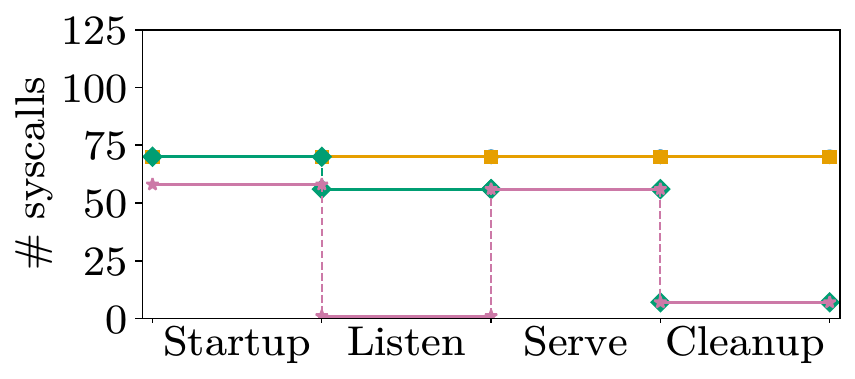}
    \caption{Lighttpd}
  \end{subfigure}
  \begin{subfigure}[b]{0.49\linewidth}
    \includegraphics[width=\linewidth]{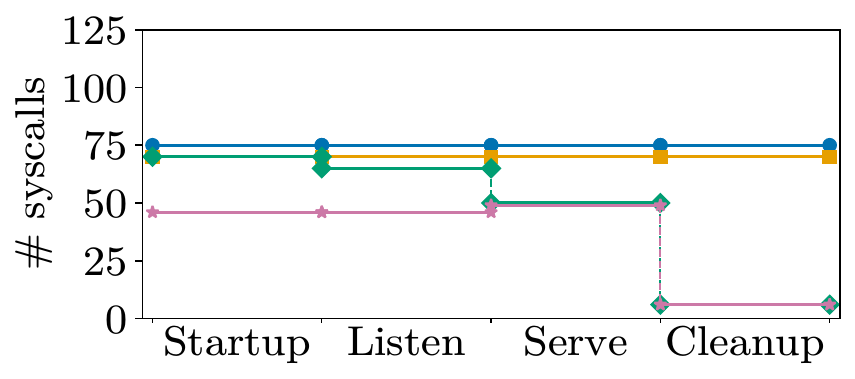}
    \caption{Nginx}
  \end{subfigure}
  \begin{subfigure}[b]{0.49\linewidth}
    \includegraphics[width=\linewidth]{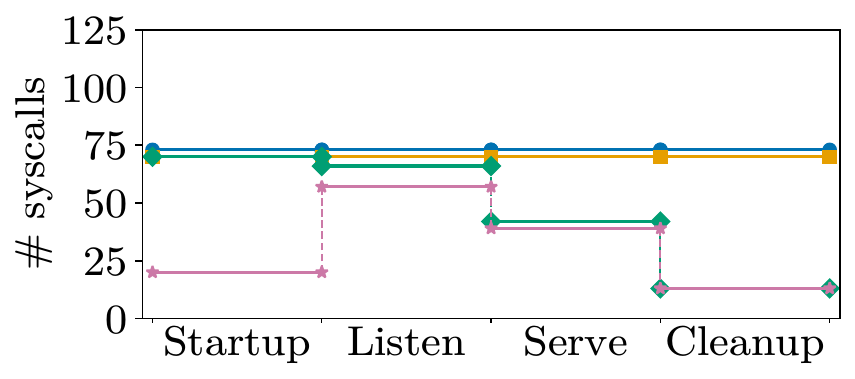}
    \caption{Memcached}
  \end{subfigure}
  \begin{subfigure}[b]{0.49\linewidth}
    \includegraphics[width=\linewidth]{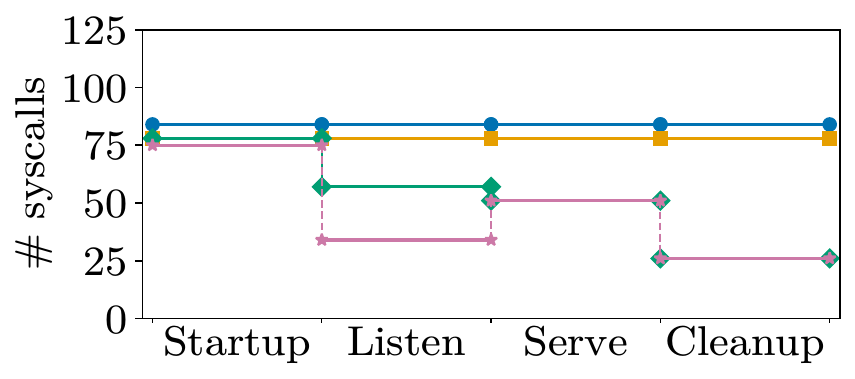}
    \caption{MongoDB}
  \end{subfigure}
  \begin{subfigure}[b]{0.49\linewidth}
    \includegraphics[width=\linewidth]{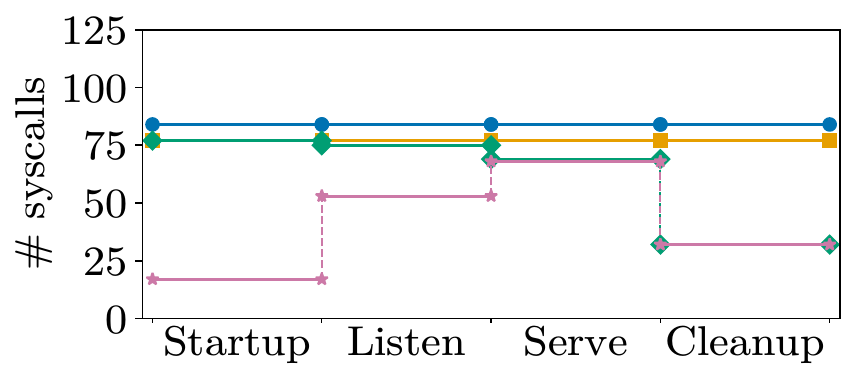}
    \caption{Redis}
  \end{subfigure}
    \begin{subfigure}[b]{0.49\linewidth}
    \includegraphics[width=\linewidth]{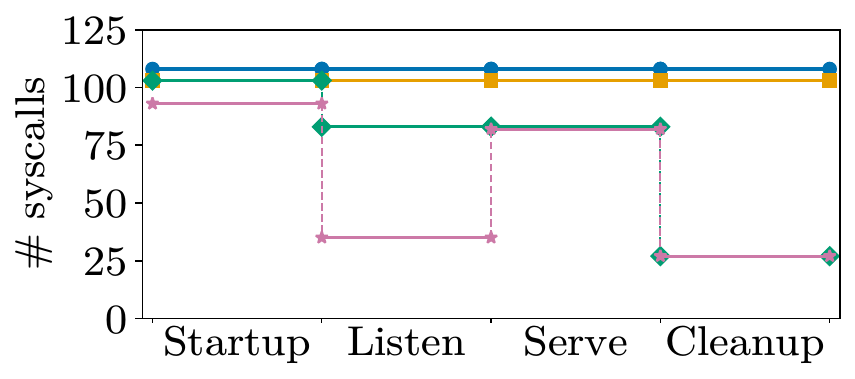}
    \caption{MySQL}
  \end{subfigure}
  \begin{subfigure}[b]{0.49\linewidth}
    \includegraphics[width=\linewidth]{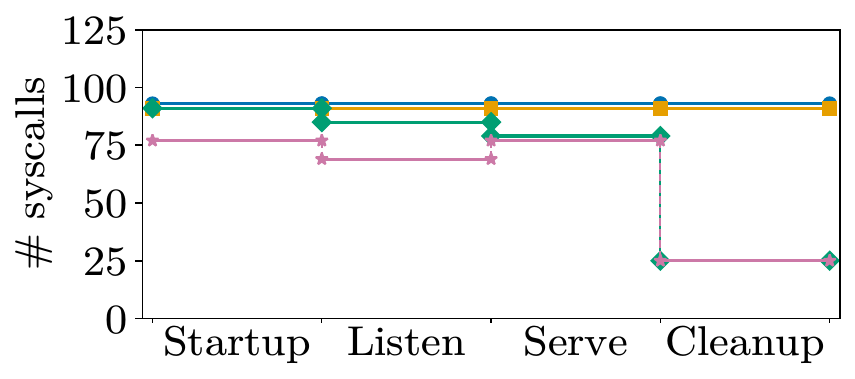}
    \caption{PostgreSQL}
  \end{subfigure}
  \begin{subfigure}[b]{0.49\linewidth}
    \includegraphics[width=\linewidth]{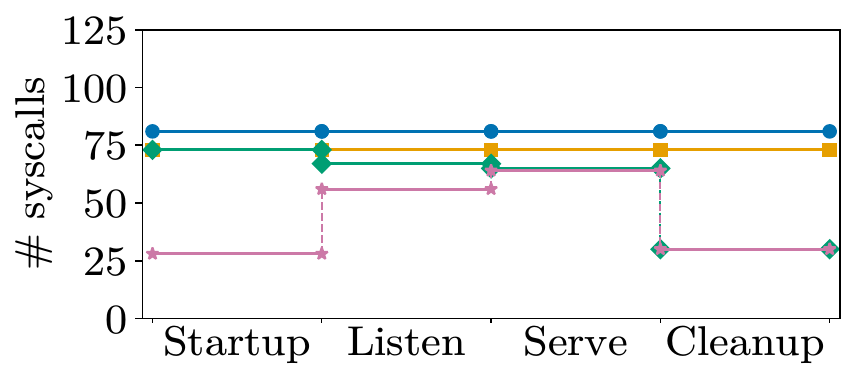}
    \caption{Bind}
  \end{subfigure}
  \caption{Number of system calls allowed in execution states of representative
    applications by strace, \syscomb without specialization ($SC_\text{mono}$), and \syscomb
    with the \emph{seccomp-like} ($SC_{\text{sec}}$) and \emph{least-privilege} ($SC_\text{lp}$)
    strategies.}\label{fig:multi-state-comparison}
\end{figure*}

We assessed the accuracy of system call detection by tracing the
execution of common server applications with both \textit{strace}~\cite{strace} and \syscomb.
Specifically, we tested Apache Httpd, Lighttpd, Nginx, Memcached, MongoDB,
Redis, MySQL, PostgreSQL and Bind by running their respective test
suites.\footnote{The test coverage averages 79.3\% across applications.}
We then measured the filter size, defined as the number of system calls
permitted. Results are shown in \cref{fig:multi-state-comparison}.
At first glance, a comparison between strace
and \syscomb without specialization ($SC_\text{mono}$) may suggest that \syscomb misses some system
calls captured by strace. However, further analysis reveals that this
discrepancy stems from the specific operational behavior of strace,
which begins tracing before the invocation of the
\texttt{execve} required to start the monitored application, thus 
capturing all system calls involved in loading and initializing the binary,
including those invoked by \texttt{libc} within \texttt{__libc_start_main}.
These system calls are recorded even if they are not required by thee
application logic, leading to overly-permissive filters.
By contrast, \syscomb begins monitoring from the application
entry point. This produces tighter system call filters,
and is aligned with existing sandboxing tools such as Minijail~\cite{minijail},
which executes \texttt{__libc_start_main} before
activating the sandbox~\cite{libminijailpreload}. To validate this,
we re-ran the applications within Minijail sandbox using the filters
identified by \syscomb. As expected, all applications
successfully completed their test suites without errors.

\subsection{Fine-Grained Specialized Filters}\label{sec:state-aware-filtering}

To evaluate the benefits of state-aware filtering, we compared the system
call filters generated by \syscomb with and without temporal specialization. Our
goal was to determine whether incorporating execution context yields
more restrictive policies. For each application mentioned in~\cref{subsect:eval-tracing}, we identified three key
transition points, partitioning task execution into four states corresponding
to major server execution phases: \emph{Startup}, \emph{Listen}, \emph{Serve}, and
\emph{Cleanup}. The states capture (i) initialization prior to
accepting requests, (ii) waiting for connection requests from the client,
(iii) serving a specific request, and (iv) termination.
\Cref{fig:multi-state-comparison} summarizes the results. Overall,
state-aware filtering consistently produces more restrictive policies.
Indeed, although the \textit{seccomp-like} strategy ($SC_{\text{sec}}$) initially
applies a filter identical to the case without specialization
($SC_\text{mono}$), it progressively transitions to stricter filters,
reducing the number of allowed system calls across all applications on
average by 11.82\% at Listen, 20.77\% at Serve, and 76.34\% at
Cleanup. The \textit{least-privilege} strategy
($SC_\text{lp}$) instead, compared to $SC_\text{mono}$, further reduces the number of system
calls by 35.22\% at Startup (which does not need to include
system calls required by subsequent states), by 47.97\% at Listen, 
22.05\% at Serve, and again 76.34\% at Cleanup.
This is also visible by comparing $SC_{\text{sec}}$ and $SC_\text{lp}$ in~\cref{fig:multi-state-comparison}, which
clearly shows that different states (with the exception of the last
one, Cleanup) rely on different subsets of system calls, highlighting
the advantages of minimal filters.

\begin{table}[t]
  \newcommand{\mrow}[2]{\multirow{#1}{*}{#2}}
  \newcommand{\LP}{\mbox{\color{Green}\ding{51}\kern-0.4em\ding{51}}}
  \newcommand{\Mono}{\ding{51}}
  \newcommand*{\All}{\Mono}
  \newcommand{\None}{\tikz[baseline=-2.0pt]\draw (0, 0) circle (0.5ex);}
  \caption{Blocked security-sensitive system calls.}\label{tab:security-sensitive-syscall}
  \addtolength{\tabcolsep}{-0.25em}
  \centering
  \begin{tabularx}{0.92\linewidth}{l *{17}{C}c}
  \toprule
  \textbf{Application}
    & \rotatebox{60}{accept}
    & \rotatebox{60}{accept4}
    & \rotatebox{60}{bind}
    & \rotatebox{60}{chmod}
    & \rotatebox{60}{clone}
    & \rotatebox{60}{connect}
    & \rotatebox{60}{execve}
    & \rotatebox{60}{execveat}
    & \rotatebox{60}{fork}
    & \rotatebox{60}{listen}
    & \rotatebox{60}{mprotect}
    & \rotatebox{60}{ptrace}
    & \rotatebox{60}{recvfrom}
    & \rotatebox{60}{setgid}
    & \rotatebox{60}{setreuid}
    & \rotatebox{60}{setuid}
    & \rotatebox{60}{socket} \\
  \midrule
  \textbf{Apache Httpd}   & \LP   & \None & \None & \None & \None & \None & \None & \All  & \Mono & \LP   & \None & \All  & \None & \All  & \All  & \All  & \None \\
  \textbf{Lighttpd}       & \LP   & \None & \LP   & \All  & \Mono & \None & \None & \All  & \All  & \LP   & \None & \All  & \None & \All  & \All  & \All  & \None \\
  \textbf{Nginx}          & \Mono & \None & \None & \None & \LP   & \None & \Mono & \All  & \All  & \LP   & \LP   & \All  & \None & \All  & \All  & \All  & \None \\
  \textbf{Memcached}      & \Mono & \None & \LP   & \All  & \LP   & \LP   & \Mono & \All  & \All  & \LP   & \None & \All  & \None & \All  & \All  & \All  & \None \\
  \textbf{MongoDB}        & \LP   & \All  & \LP   & \LP   & \Mono & \LP   & \Mono & \All  & \All  & \LP   & \None & \All  & \All  & \All  & \All  & \All  & \LP   \\
  \textbf{Redis}          & \Mono & \None & \None & \Mono & \None & \None & \Mono & \All  & \All  & \None & \None & \All  & \None & \All  & \All  & \All  & \None \\
  \textbf{MySQL}          & \None & \All  & \None & \LP   & \LP   & \None & \Mono & \All  & \All  & \None & \None & \All  & \None & \All  & \All  & \All  & \None \\
  \textbf{PostgreSQL}     & \LP   & \All  & \LP   & \None & \None & \None & \None & \All  & \All  & \LP   & \None & \All  & \None & \All  & \All  & \All  & \None \\
  \textbf{Bind}           & \Mono & \None & \None & \Mono & \Mono & \None & \Mono & \All  & \All  & \None & \None & \All  & \All  & \All  & \All  & \All  & \None \\
  \bottomrule
  \multicolumn{18}{l}{Filtered \LP~with specialization, \Mono~without specialization, and \None~by no method}
  \end{tabularx}
\end{table}

\begin{figure}[t]
  \centering
  \includegraphics[width=\linewidth]{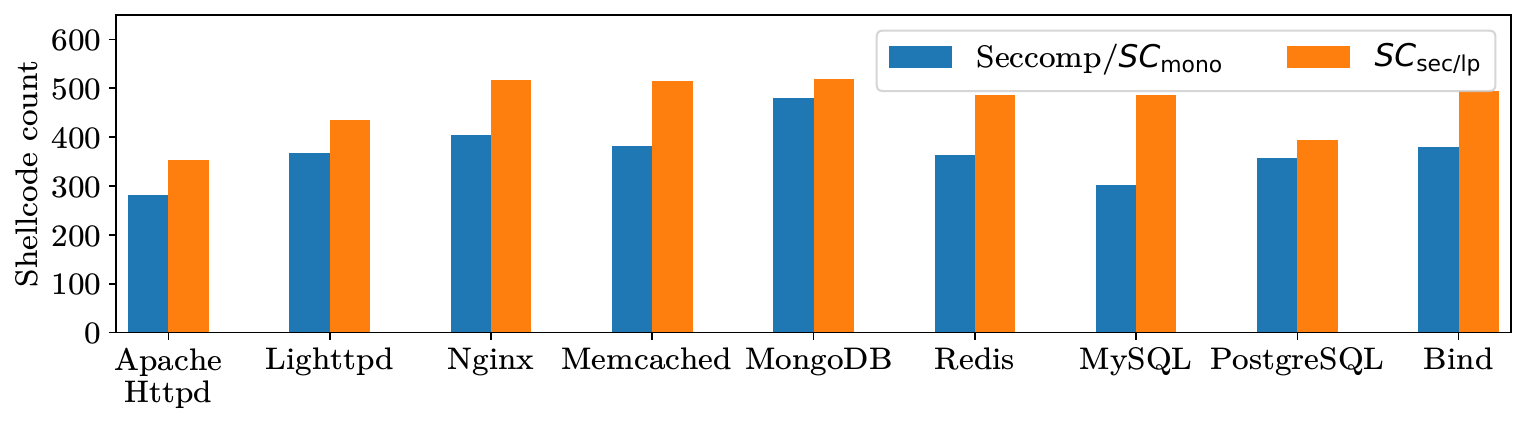}
  \caption{Shellcodes blocked over a total of 535 shellcodes.}\label{fig:shellcodes-stopped}
\end{figure}

\subsection{Attack Surface Reduction}\label{sec:surface-reduction}

To showcase the attack surface reduction, we focused on the ability to
block kernel vulnerabilities.  To this end, we first selected a sample
of 45 recent kernel CVEs (shown in~\cref{tab:kernel-cves} of
Appendix~\ref{appendix:cve}) that are exploited through specific
system calls. Then, we re-examined the applications discussed
in~\Cref{subsect:eval-tracing}, and counted how many of them are
protected when
\begin{enumerate*}
\item a single filter is used, set by either seccomp or
  \syscomb without specialization,
and when
\item a specialized fiter is used for the Serve state. 
\end{enumerate*}
The data show that a specialized filter leads to better protection
in 17 out of 45 cases (37.8\% of CVEs).
This result originates from the ability to more effectively filter
\emph{security-sensitive} system calls, defined
in~\cite{temporal-specialization} as those frequently leveraged in
shellcodes. Indeed, across the 9 tested applications, the 17
security-sensitive system calls reported in
\cref{tab:security-sensitive-syscall} are blocked only in 55.9\% of
cases by \syscomb without specialization, and in 72.8\% cases when
specialization is used (+16.9\%).
With reference to the set of 535 shellcodes identified by previous
research~\cite{temporal-specialization,syspart}, this translates to
blocking on average 98 more shellcodes per application (+27.8\%), as
shown in \cref{fig:shellcodes-stopped}.

We remind that, while \syscomb provides a unique solution to both
identify and enforce specialized filters, it can also enforce the filters
generated by other
approaches~\cite{temporal-specialization,syspart,sysfilter}. Moreover,
it removes the need to inject calls to seccomp for the activation, it
can work with more than two states, and it supports least-privilege
filters. No user-space monitors and kernel modifications are required.

\section{Performance Evaluation}\label{sec:runtime-evaluation}

To assess \syscomb's performance we employed
\begin{enumerate*}
  \item microbenchmarks, to measure the absolute system
    call interposition overhead, and
  \item macrobenchmarks, to evaluate the impact on common software.
\end{enumerate*}
Tests were run on a server with Linux kernel~6.8.0, Threadripper
2950X CPU, 128~GiB RAM, and 1~TiB~SSD.\@

\paragraph{Microbenchmarks}
We compared different system call interposition approaches measuring the
execution time of the \texttt{getpid} system call, known for its minimal
kernel execution path and thus ideal to isolate the overhead. To increase the
accuracy, we employed high-resolution timers, executed 1 million consecutive
\texttt{getpid} calls, and finally calculated the average.
The test was repeated for \texttt{ptrace}~\cite{ptrace},
\texttt{LD_PRELOAD}~\cite{syscall-intercept}, interrupt-based instrumentation
(\texttt{int3}), SUD~\cite{sud}, zpoline~\cite{zpoline}, seccomp~\cite{seccomp},
and \syscomb.
Results are shown in \cref{tab:microbenchmark}.

\begin{table}[t]
  \caption{Average execution time of \texttt{getpid} over 1 million
    invocations with different interposition solutions.}\label{tab:microbenchmark}
    \centering
    \begin{tabularx}{\columnwidth}{l R R}
    \toprule
    \textbf{Interposition mechanism} & \textbf{Tracing ($ns$)} & \textbf{Enforcing ($ns$)} \\
    \midrule
    \texttt{ptrace}                  & 11931                   & 12019 \\
    \texttt{LD_PRELOAD}              & 200                     & 201 \\
    \texttt{int3}                    & 2880                    & 3157 \\
    SUD                              & 2658                    & 2731 \\
    zpoline                          & 198                     & 199 \\
    seccomp                          & 5167                    & 217 \\
    \syscomb~(Raw tracepoint)        & 297                     & 303 \\
    \syscomb~(Kprobes)               & n.a.                    & 487 \\
    \bottomrule
    \multicolumn{3}{l}{Native execution (without interposition) has an average execution time of 166 $ns$} \\
    \end{tabularx}
\end{table}

As expected, all interposition techniques introduce some overhead compared to
the native execution, which takes 166~$ns$ on average.
The solutions that require a context-switch to a user-space monitor
(\texttt{ptrace}, interrupt-based instrumentation, SUD, and seccomp in
user-space notification mode) introduce the highest overhead, with execution
times ranging from 2658~$ns$ to 12019~$ns$ per system call on average (a
slowdown of one to two orders of magnitude).
Instead, techniques that operate entirely in user-space  (\texttt{LD_PRELOAD},
zpoline) or entirely in kernel-space (seccomp in enforcing mode, \syscomb) add
lower overhead, ranging from 198~$ns$ to 487~$ns$.
Despite being performant, user-space solutions are easily bypassed (see
Section~\ref{sec:ext-solutions}), while kernel-based ones provide robust
protection.
Overall, \syscomb exhibits competitive performance, with an average execution
time of 297~$ns$ in tracing mode, 303~$ns$ in enforcement mode with raw
tracepoints, and 487~$ns$ in enforcement mode with kprobes.

\paragraph{Macrobenchmarks}
In real scenarios, applications do not spend most of their lifetime invoking
system calls, so the interposition overhead is amortized by the user-space and
kernel-space execution time.
To assess this, we ran several macrobenchmarks, measuring the native execution
time (the baseline) and the execution time under seccomp and \syscomb
interposition. The tested applications include Apache Httpd, Lighttpd, Nginx,
Memcached, MongoDB, Redis, MySQL, PostgreSQL, and Bind. For benchmarking, we
used
\begin{enumerate*}
  \item Wrk~\cite{wrk} to send HTTP requests,
  \item YCSB~\cite{ycsb} to submit key-value read and write queries,
  \item SysBench~\cite{sysbench} and \texttt{pgbench}~\cite{pgbench} to execute
        SQL queries,
  and
  \item \texttt{dnsperf}~\cite{dnsperf} to resolve DNS names.
\end{enumerate*}
\cref{tab:macrobenchmark} shows the 90th percentiles of request latency.

Overall, in tracing mode, compared to the baseline \syscomb adds a minimum of 0.002~$ms$ (Memcached)
and a maximum of 14.596~$ms$ (Bind), with the latter
accounting for only 4.46\% of Bind's request time. In enforcement mode,
\syscomb's overhead ranges from 0.000~$ms$ (MySQL with raw tracepoints) to 2.233~$ms$
(Bind with kprobes) compared to seccomp, with the latter accounting for only
0.68\% of Bind's request time. The results demonstrate low overhead,
suggesting the viability of our approach in production environments.

To conclude, we highlight the impact of the enforcement action.
When errors are injected and returned to the application, the overhead is
slightly higher. This is caused by the use of kprobes over raw tracepoints,
which require a different eBPF program type with dedicated safety checks, and
additional steps to retrieve the system call number (e.g., cookie-based lookup).
Nevertheless, the impact remains limited, with a maximum measured latency of
$0.521~ms$ (MySQL).

\newcommand{\hdrl}[2]{\multicolumn{#1}{l}{\textbf{#2}}}
\newcommand{\hdrc}[2]{\multicolumn{#1}{c}{\textbf{#2}}}
\newcommand{\hdrr}[2]{\multicolumn{#1}{r}{\textbf{#2}}}
\begin{table}[t]
  \caption{90th percentile response time with different interposition techniques. All
  values are in milliseconds.}\label{tab:macrobenchmark}
  \centering
  \begin{tabularx}{\columnwidth}{l *{5}{R}}
  \toprule
    \hdrl{1}{\multirow{2}{*}{Application}}
    & \hdrc{1}{\multirow{2}{*}{Baseline}}
    & \hdrc{1}{\multirow{2}{*}{seccomp}}
    & \hdrc{1}{\multirow{2}{*}{\syscomb (tr.)}}  & \hdrc{2}{\syscomb (enf.)}   \\
    & & & & \hdrc{1}{Raw T.} & \hdrc{1}{Kprobes} \\
  \midrule
  Apache Httpd  & 7.731    & 7.743   & 8.317   & 8.327   & 8.531    \\
  Lighttpd      & 11.442   & 12.578  & 12.631  & 12.790  & 12.852   \\
  Nginx         & 7.407    & 7.414   & 8.138   & 8.169   & 8.411    \\
  Memcached     & 0.075    & 0.077   & 0.077   & 0.078   & 0.080    \\
  MongoDB & 0.163    & 0.166   & 0.169   & 0.170   & 0.169    \\
  Redis         & 0.104    & 0.106   & 0.108   & 0.109   & 0.108    \\
  MySQL         & 27.165   & 28.673  & 28.673  & 28.673  & 29.194   \\
  PostgreSQL    & 5.034    & 5.101   & 5.111   & 5.141   & 5.107    \\
  Bind          & 312.776  & 326.209 & 327.372 & 328.116 & 328.442  \\
  \bottomrule
  \end{tabularx}
\end{table}
\let\hdrc\undefined

\section{Conclusions}\label{sec:conclusions}

We presented \syscomb, a novel solution to support the activation of specialized
system call filters. Unlike existing work, \syscomb requires no modification of
application and kernel code, and introduces no user-space monitors. Indeed, by
leveraging eBPF, \syscomb operates entirely within the kernel to automatically
track application behavior and enable the fully transparent, synchronous,
non-bypassable enforcement of per-thread system call filters for any
application state. We implemented and evaluated \syscomb on nine widely deployed
applications, including web servers, relational and NoSQL DBMSs, and a DNS
resolver. The results show accurate system call detection and increased filter
selectivity, leading to improved protection against kernel vulnerabilities
and exploit payloads. Standard benchmarks indicate comparable performance to
seccomp. Overall, \syscomb offers an effective, low-overhead solution that
reduces the kernel attack surface and robustly mitigates attacks in real-world
applications.

\bibliographystyle{splncs04}
\bibliography{bibliography.bib}

\newpage
\appendix
\section{Backpropagation Algorithms}\label{sec:backpropagation}

We provide a high-level representation of the algorithms:
\cref{alg:flow-backpropagation} details the flow-based backpropagation
of system calls, \cref{alg:strongly-connected-components} reports the
Kosaraju-Sharir's algorithm to find the the strongly connected
components, \cref{alg:dfs} performs a Depth-First Search visit, and
\cref{alg:transpose-graph} transposes the graph.
\vspace{-1.5em}
\begin{algorithm}
\caption{Flow-based backpropagation of system call profiles.}\label{alg:flow-backpropagation}
\begin{algorithmic}[1]
\Procedure{Backpropagation}{$states$}
\State $sccs \gets$ \Call{StronglyConnectedComponents}{$states$}
\ForAll{$scc$ in \Call{Reverse}{$sccs$}}
    \State $syscalls \gets \emptyset$
    \ForAll{$state$ in $scc$}
        \State $syscalls \gets syscalls \cup state.syscalls$
    \EndFor
    \ForAll{$state$ in $scc$}
        \State $state.syscalls \gets syscalls$
        \ForAll{$prev$ in $state.prev$}
            \State $prev.syscalls \gets prev.syscalls \cup state.syscalls$
        \EndFor
    \EndFor
\EndFor
\EndProcedure
\end{algorithmic}
\end{algorithm}
\vspace{-1.5em}
\begin{algorithm}[h!]
\caption{Identification of strongly connected components.}\label{alg:strongly-connected-components}
\begin{algorithmic}[1]
\Procedure{StronglyConnectedComponents}{$states$}
\State $visited \gets \emptyset$
\State $stack \gets \emptyset$
\ForAll{$state$ in $states$}
    \If{$state \notin visited$}
        \State \Call{DFS}{$states$, $state$, $visited$, $stack$}
    \EndIf
\EndFor
\State $transposed \gets$ \Call{TransposeGraph}{$states$}
\State $visited \gets \emptyset$
\State $sccs \gets \emptyset$
\While{$stack \neq \emptyset$}
    \State $state \gets stack.pop()$
    \If{$state \notin visited$}
        \State $component \gets \emptyset$
        \State \Call{DFS}{$transposed$, $state$, $visited$, $component$}
        \State $sccs.append(component)$
    \EndIf
\EndWhile
\State \Return $sccs$
\EndProcedure
\end{algorithmic}
\end{algorithm}
\begin{algorithm}
\caption{DFS from a given state.}\label{alg:dfs}
\begin{algorithmic}[1]
\Procedure{DFS}{$states$, $state$, $visited$, $stack$}
\State $visited \gets visited \cup \{state\}$
\ForAll{$next$ in $states[state].next$}
    \If{$next \notin visited$}
        \State \Call{DFS}{$states$, $next$, $visited$, $stack$}
    \EndIf
\EndFor
\State $stack.push(state)$
\EndProcedure
\end{algorithmic}
\end{algorithm}
\begin{algorithm}
\caption{Graph transposition.}\label{alg:transpose-graph}
\begin{algorithmic}[1]
\Procedure{TransposeGraph}{$states$}
\State $trans \gets$ new graph
\ForAll{$state$ in $states$}
    \ForAll{$next$ in $states[state].next$}
        \State $trans[next].next \gets trans[next].next \cup \{state\}$
    \EndFor
\EndFor
\State \Return $trans$
\EndProcedure
\end{algorithmic}
\end{algorithm}

\newpage
\section{Kernel CVEs Blocked}\label{appendix:cve}

\begin{table}[h!]
  \newcommand{\mrow}[2]{\multirow{#1}{*}{#2}}
  \caption{Number of application-specific system call filters identified by
    \syscomb that successfully mitigates Kernel CVEs using a single or
    a specialized filter. Further details
    in~\cref{sec:surface-reduction}.}\label{tab:kernel-cves}
  \scriptsize
  \addtolength{\tabcolsep}{0.8em}
  \centering
  \begin{tabular}{ll cc}
    \toprule
\textbf{CVE}                    & \textbf{System calls}      & \textbf{$SC_\text{mono}$}  & \textbf{$SC_\text{sec/lp}$}     \\
\midrule
2010-4243                       & uselib, execve(at)         & \textbf{6}           & \textbf{6}             \\
\midrule
\mrow{2}{2011-1082}             & epoll_ctl, epoll_pwait     & \mrow{2}{0}          & \mrow{2}{\textbf{2}}   \\
                                & epoll_wait \\
\midrule
2012-3375                       & epoll_ctl                  & 0                    & \textbf{2}             \\
\midrule
2013-1959                       & write                      & 0                    & 0                      \\
\midrule
2013-1979                       & recvfrom, recvmsg          & 0                    & \textbf{1}             \\
\midrule
3014-4699                       & fork, clone, ptrace        & 3                    & \textbf{6}             \\
\midrule
2014-5207                       & mount                      & \textbf{9}           & \textbf{9}             \\
\midrule
2014-7970                       & pivot_root                 & \textbf{9}           & \textbf{9}             \\
\midrule
2015-8543                       & socket                     & 0                    & \textbf{1}             \\
\midrule
2014-9529, 2016-0728            & keyctl                     & \textbf{9}           & \textbf{9}             \\
\midrule
2016-2383                       & bpf                        & \textbf{9}           & \textbf{9}             \\
\midrule
2017-6001                       & perf_event_open            & \textbf{9}           & \textbf{9}             \\
\midrule
2017-14954                      & waitid                     & \textbf{9}           & \textbf{9}             \\
\midrule
2017-17712                      & sendto, sendmsg            & 0                    & \textbf{2}             \\
\midrule
2017-18509                      & setsockopt, getsockopt     & 0                    & \textbf{2}             \\
\midrule
2008-3527, 2010-3858, 2010-4346 & \mrow{4}{execve(at)}       & \mrow{4}{\textbf{6}} & \mrow{4}{\textbf{6}}   \\
2012-4530, 2013-0914, 2014-9585 &                            &                      &                        \\
2015-1593, 2015-3339, 2016-3672 &                            &                      &                        \\
2018-14634                      &                            &                      &                        \\
\midrule
2018-11508                      & adjtimex                   & \textbf{9}           & \textbf{9}             \\
\midrule
2018-12233                      & setxattr                   & \textbf{9}           & \textbf{9}             \\
\midrule
2018-18281                      & execve(at), mremap         & \textbf{5}           & \textbf{5}             \\
\midrule
2019-10125                      & io_submit                  & \textbf{8}           & \textbf{8}             \\
\midrule
2019-11815                      & clone, unshare             & 3                    & \textbf{6}             \\
\midrule
2018-1000199, 2019-13272        & ptrace                     & \textbf{9}           & \textbf{9}             \\
\midrule
2016-4998, 2016-4997            & \mrow{2}{setsockopt}       & \mrow{2}{1}          & \mrow{2}{\textbf{3}}   \\
2016-3134, 2021-32606           &                            &                      &                        \\
\midrule
2021-3609                       & sendmsg                    & \textbf{5}           & \textbf{6}             \\
\midrule
2021-44879                      & openat2                    & \textbf{9}           & \textbf{9}             \\
\midrule
2022-1198                       & sendmsg, bind              & 0                    & \textbf{3}             \\
\midrule
2022-1204                       & connect, bind              & 0                    & \textbf{2}             \\
\midrule
2022-47929                      & sendto                     & 0                    & \textbf{3}             \\
\midrule
2022-47943                      & preadv2                    & \textbf{9}           & \textbf{9}             \\
\midrule
2023-23454                      & connect, sendto            & 0                    & \textbf{2}             \\
\midrule
2023-25012                      & bind                       & 0                    & \textbf{4}             \\
    \bottomrule
  \end{tabular}
\end{table}


\end{document}